\documentclass[prd,twocolumn]{revtex4}
\usepackage{dcolumn}
\usepackage{multirow}
\usepackage{graphicx}
\usepackage{amssymb}
\usepackage{bm}
\usepackage{hyperref}
\usepackage{epstopdf}
\usepackage{color}
\usepackage{mathrsfs}
\usepackage{amsmath,amssymb,amsthm}
\usepackage{rotating}
\usepackage{sverb, longtable}
\usepackage{subfigure}

\usepackage[graphicx]{realboxes}
\usepackage{adjustbox}

\begin{document}

\title{The Self-Consistency of DESI Analysis and Comment on ``Does DESI 2024 Confirm $\Lambda$CDM?''}

\author{Deng Wang}
\email{dengwang@ific.uv.es}
\affiliation{Instituto de F\'{i}sica Corpuscular (CSIC-Universitat de Val\`{e}ncia), E-46980 Paterna, Spain}

\begin{abstract}
We demonstrate that the constraints on the evolution of dark energy implemented by the DESI collaboration may be insufficient or incomplete using their own BAO data. Using large enough prior ranges for the present-day equation of state of dark energy $\omega_0$ and amplitude of dark energy evolution $\omega_a$, we obtain the complete $1\,\sigma$ and $2\,\sigma$ constraints $\omega_0=1.04^{+0.91+2.00}_{-1.00-1.90}$ and $\omega_a=-7.4^{+3.8+6.8}_{-3.2-7.3}$ indicating a beyond $2\,\sigma$ preference of quintessence-like dark energy today and an evidence of evolving dark energy at beyond $2\,\sigma$ CL, respectively. Our results are different from $\omega_0=-0.55^{+0.39}_{-0.21}$ and the $2\,\sigma$ upper limit $\omega_a<-1.32$ reported by the DESI collaboration \cite{DESI:2024mwx}. Employing a data combination of cosmic microwave background, DESI BAO and type Ia supernova, we obtain the $1\,\sigma$, $2\,\sigma$ and $3\,\sigma$ constraints $\omega_0=-0.707^{+0.089+0.18+0.24}_{-0.089-0.17-0.22}$ and $\omega_a=-1.09^{+0.38+0.67+0.82}_{-0.31-0.72-1.00}$, which reveals a $\sim4\,\sigma$ evidence of dynamical dark energy when the redshift $z\lesssim0.1$. We verify that the BAO data point from luminous red galaxies at the effective redshift $z_{\rm eff}=0.51$ hardly affects the joint constraint from the data combination of cosmic microwave background, DESI BAO and type Ia supernova. We also point out the shortcomings and advantages of the binning method widely used in cosmological analyses.

\end{abstract}
\maketitle

\section{Introduction}
$\Lambda$CDM is expected to depict the universe at both large and small scales through the whole cosmic history. However, it has faced various kinds of problems and tensions emerged during the past two decades \cite{DiValentino:2020vhf,Abdalla:2022yfr,DiValentino:2020zio}. Recently, the DESI collaboration release the first year of observations from the Dark Energy Spectroscopic Instrument (DESI), in which they report new high-precision BAO measurements \cite{DESI:2024uvr,DESI:2024lzq} from galaxy, quasar and Lyman-$\alpha$ forest tracers. The corresponding cosmological analysis from these BAO datasets are performed in Ref.\cite{DESI:2024mwx}, where they claim a $2.6\,\sigma$ preference of dark energy evolution via the data combination of cosmic microwave background (CMB) plus DESI. Interestingly, the addition of Pantheon+ \cite{Scolnic:2021amr}, Union3 \cite{Rubin:2023ovl} and DESY5 \cite{DES:2024tys} supernova (SN) increase the preference to $2.5\,\sigma$, $3.5\,\sigma$ and $3.9\,\sigma$, respectively. 

Most recently, the authors in Ref.\cite{Colgain:2024xqj} question the validity of DESI's finding of evolving dark energy by reanalyzing the DESI BAO data points. They mainly study the impact of the anomalous BAO data point at the effective redshift $z_{\rm eff}=0.51$ from luminous red galaxies (LRG) in the redshift range $0.4<z<0.6$ on the cosmological analysis and outcomes. They also derive the matter fraction $\Omega_m(z_{\rm eff}=0.51)=0.668^{+0.180}_{-0.169}$ and claim it is inconsistent with that from Planck CMB data at the $\sim2\sigma$ confidence level (CL). Nonetheless, these results has been presented or implicitly revealed by the DESI collaboration in their Figure 2 or Section 5 \cite{DESI:2024mwx}. Furthermore, they claim that the DESI's evidence of dynamical dark energy (DDE) are only believable if baryon acoustic oscillations (BAO) and SN show consistent deviations from $\Lambda$CDM in similar redshift ranges. This point of view is fairly reasonable. For example, our constraints $\Omega_m=0.43^{+0.16}_{-0.22}$ and $\Omega_m=0.50^{+0.13}_{-0.41}$ in the range $z\in[0.45, 0.68]$ is well consistent with the DESI's value at $z_{\rm eff}=0.51$ within $1\,\sigma$ CL when using the Pantheon+ SN sample \cite{Scolnic:2021amr} with and without the Cepheid host distance calibration, respectively (see Tabs. I and II in Ref.\cite{Wang:2022ssr} for details). This implies that Pantheon+ supports the large matter density ratio from DESI around $z_{\rm eff}=0.51$, and consequently helps confirm the evidence of DDE. However, the true evidence of DDE should be from the combination of CMB+BAO+SN or at least CMB+BAO. The DESI collaboration have conducted these analyses and given the substantial evidence of DDE. Basically, we think the cosmological analyses from the DESI collaboration is reasonable and self-consistent. However, the DDE constraints implemented by the DESI collaboration may be insufficient or incomplete using their own BAO data. In this short study, we aim at analyzing further the DESI data, demonstrating the self-consistency of DESI's global fit when combining DESI with CMB and SN data, and suggesting the direction of future researches.

This work is organized as follows. In the next section, we review briefly the basic formula for the $\Lambda$CDM and DDE models. In Section III, we describe the data and analysis methodology. In Section IV, we analyze the compatibility of CMB, DESI and Pantheon+ under the assumption of $\Lambda$CDM. In Section V, we verify the consistency of these three probes when constraining the Chevallier-Polarski-Linder (CPL) DDE model \cite{Chevallier:2000qy,Linder:2002et}. In Section VI, we discuss the shortcomings and advantages of the commonly used binning method. Discussions are presented in the final section.  

\section{Models}
In this section, we introduce two cosmological models to be constrained by observations. The homogeneous and isotropic universe described by the Friedmann-Lema\^{i}tre-Robertson-Walker metric (FLRW) metric
\begin{equation}
\mathrm{d}s^2=-dt^2+a^2(t)\left[\frac{\mathrm{d}r^2}{1-Kr^2}+r^2\mathrm{d}\theta^2+r^2sin^2\theta \mathrm{d}\phi^2\right],      \label{eq:FRW}
\end{equation}  
where $a(t)$ and $K$ are the scale factor at cosmic time $t$ and the Gaussian curvature of spacetime, respectively. Substituting Eq.(\ref{eq:FRW}) into the Einstein's field equation, we obtain the so-called Friedmann equations as follows
\begin{equation}
H^2=\frac{8\pi G}{3}\Sigma\rho_i,     \label{eq:f1}
\end{equation}   
\begin{equation}
\frac{\ddot{a}}{a}=-\frac{4\pi G}{3}\Sigma(\rho_i+3p_i),     \label{eq:f2}
\end{equation}   
where $H$ is the Hubble parameter, and $\rho_i$ and $p_i$ represent the mean energy density and pressure of different species in the cosmic pie. Combining Eqs.(\ref{eq:f1}) with (\ref{eq:f2}), one can easily obtain the dimensionless Hubble parameter (DHP), which depicts the background evolution of a specific cosmological model, for the $\Lambda$CDM scenario
\begin{equation}
E_{\mathrm{\Lambda CDM}}(z)=\left[\Omega_{m}(1+z)^3+1-\Omega_{m}\right]^{\frac{1}{2}}. \label{eq:ezlcdm}
\end{equation}

Furthermore, to study the dark energy evolution, we consider the so-called CPL parameterization \cite{Chevallier:2000qy,Linder:2002et}. Its DHP reads as 
\begin{widetext}
\begin{equation}
E_{\mathrm{CPL}}(z)=\left[\Omega_{m}(1+z)^3+(1-\Omega_{m})(1+z)^{3(1+\omega_0+\omega_a)\mathrm{e}^{\frac{-3\omega_az}{1+z}}}\right]^{\frac{1}{2}}, \label{eq:ezcpl}
\end{equation}
\end{widetext}
which reduces to $\Lambda$CDM when $\omega_0=-1$ and $\omega_a=0$.

\begin{table}[!t]
	\renewcommand\arraystretch{1.6}
	\begin{center}
		\caption{Mean values and $1\,\sigma$ ($2\,\sigma$) uncertainties of the parameter pair ($\omega_0$, $\omega_a$) in the CPL model from the DESI BAO measurements with and without the LRG1 data point.}
		\setlength{\tabcolsep}{2mm}{
			\label{tab:CPL}
			\begin{tabular}{c|c|c}
				\hline
				\hline
				Parameters & DESI & DESI without LRG1\\
				\hline 
				$\omega_0$&  $1.04^{+0.91+2.00}_{-1.00-1.90}$  & $1.0^{+1.4+3.7}_{-2.0-3.1}$  \\
				\hline
				$\omega_a$&  $-7.4^{+3.8+6.8}_{-3.2-7.3}$   &  $-7.5^{+7.0+11.0}_{-4.8-13.0}$ \\
				\hline
				\hline
		\end{tabular}}
	\end{center}
\end{table}

\begin{figure}
	\centering
	\includegraphics[scale=0.5]{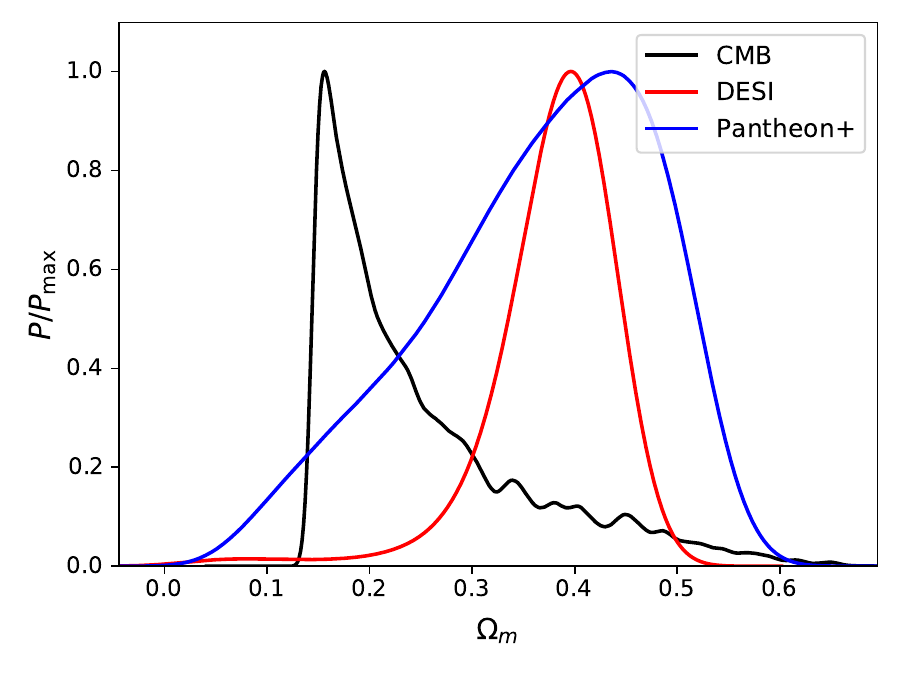}
	\caption{One-dimensional posterior distributions of the matter density ratio $\Omega_m$ from CMB, DESI and SN observations in the CPL model.}\label{fig:CPLom}
\end{figure}

\begin{figure*}
	\centering
	\includegraphics[scale=0.6]{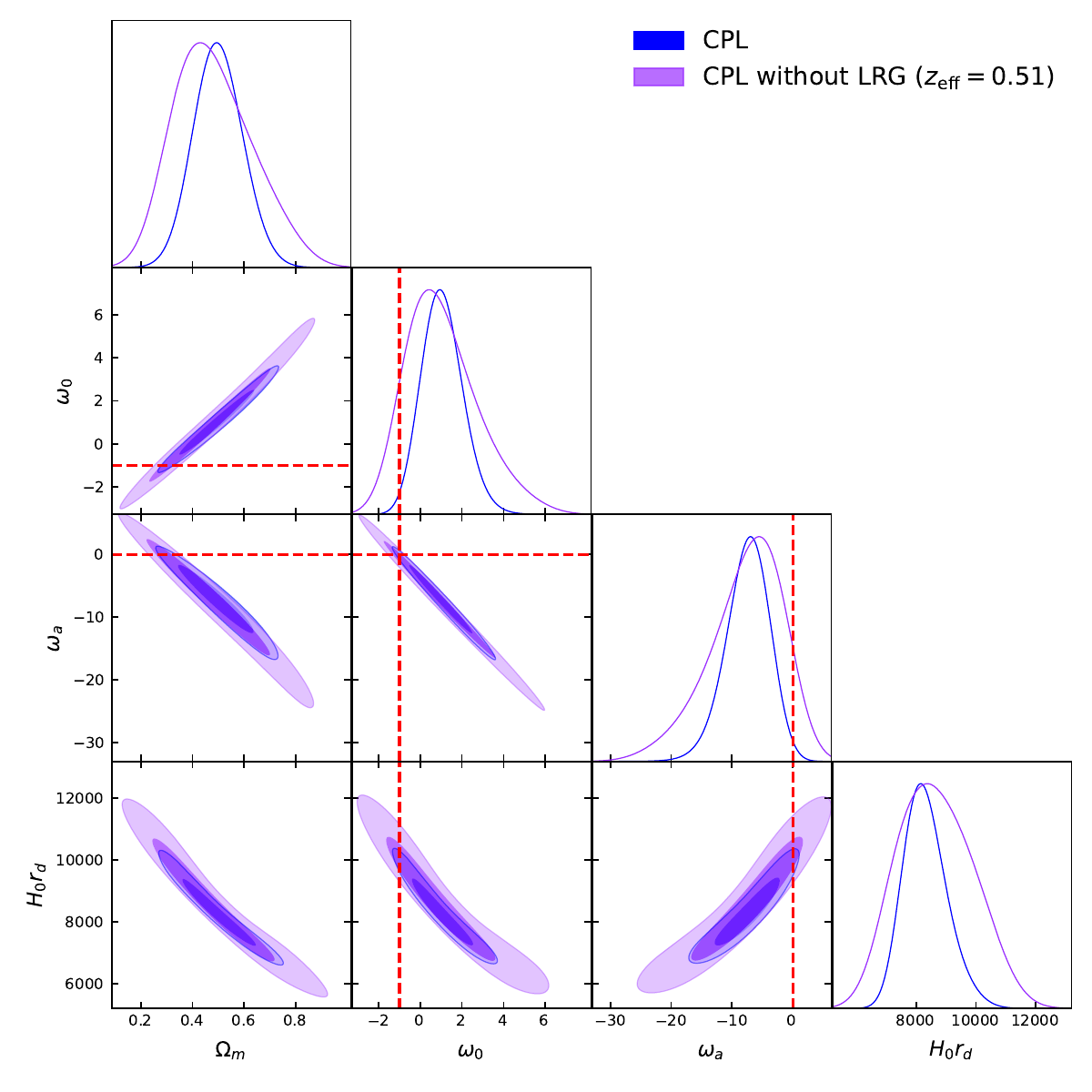}
	\caption{One-dimensional and two-dimensional posterior distributions of free parameters in the CPL model from the DESI BAO data with and without the LRG1 data point. The red dashed lines denote $\omega_0=-1$ (or $\omega_a=0$). The cross point between dashed lines is the $\Lambda$CDM scenario.}\label{fig:CPLLRG}
\end{figure*}


\begin{figure}
	\centering
	\includegraphics[scale=0.5]{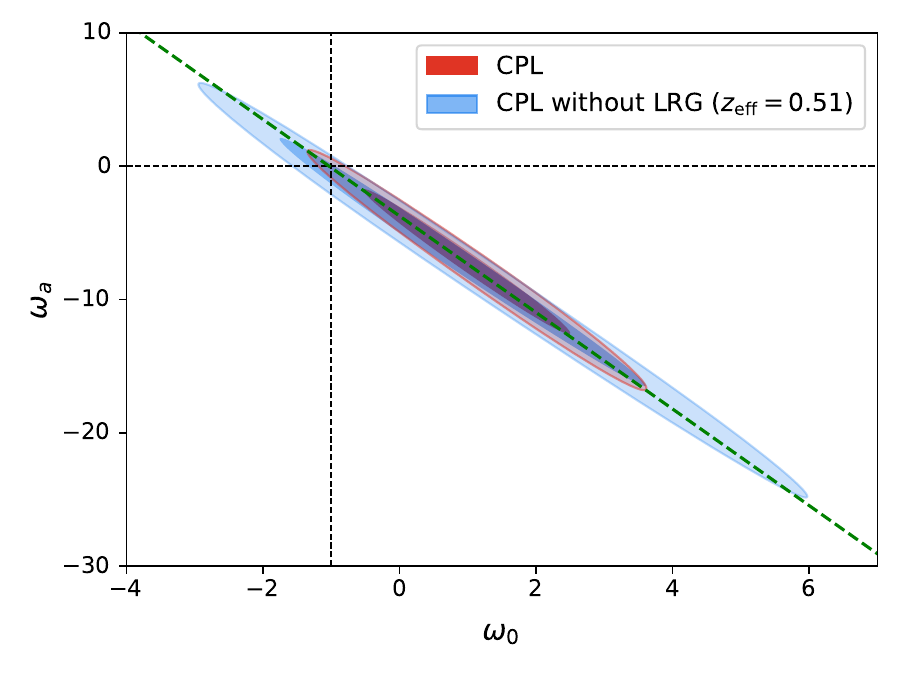}
	\caption{Two-dimensional posterior distributions of the parameter pair ($\omega_0$, $\omega_a$) in the CPL model from the DESI BAO measurements with and without the LRG1 data point. The cross point between the black dashed lines is the $\Lambda$CDM scenario. The green dashed line denotes the approximately linear combination $3.62\times\omega_0+\omega_a=-3.73$. Here $r_d$ is the comoving sound horizon at the drag epoch.}\label{fig:CPLlinear}
\end{figure}

\begin{figure}
	\centering
	\includegraphics[scale=0.5]{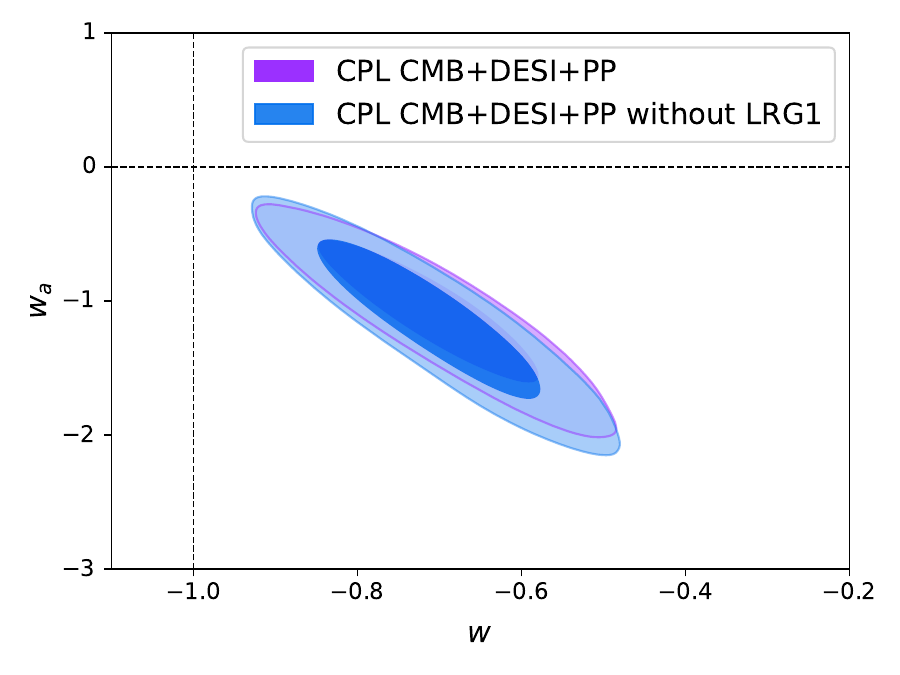}
	\caption{Two-dimensional posterior distributions of the parameter pair ($\omega_0$, $\omega_a$) in the CPL model from the data combination of Planck, DESI and Pantheon+ with and without the LRG1 data point. The cross point between the black dashed lines is the $\Lambda$CDM scenario.}\label{fig:CPLCDS}
\end{figure}

\begin{figure*}
	\centering
	\includegraphics[scale=0.5]{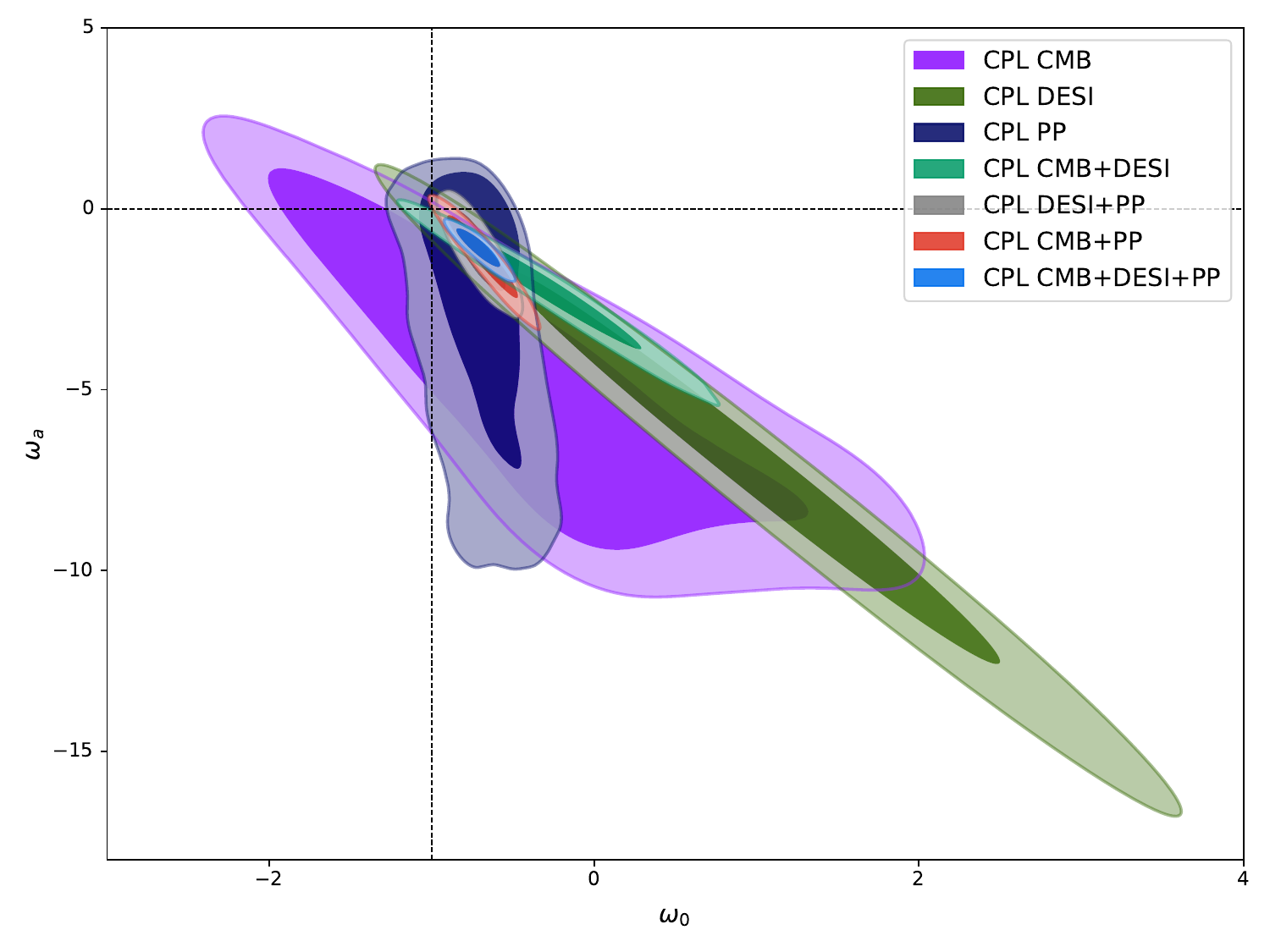}
	\caption{Two-dimensional posterior distributions of the parameter pair ($\omega_0$, $\omega_a$) from different probes and their combinations in the CPL model. Here PP is short for Pantheon+. The cross point between the black dashed lines is the $\Lambda$CDM scenario.}\label{fig:CPLw0wa}
\end{figure*}

\begin{figure}
	\centering
	\includegraphics[scale=0.55]{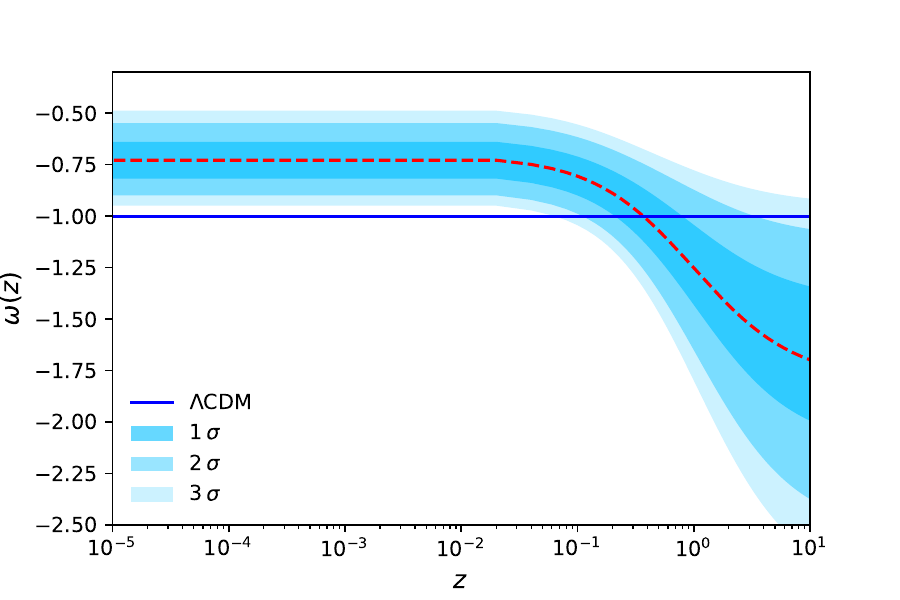}
	\caption{The constrained dark energy parameter space in the CPL model from the data combination of CMB, DESI and SN. The red dashed and blue solid lines denote the best fit cosmology and the $\Lambda$CDM model, respectively. The shaded regions are the $1\sigma$, $2\sigma$ and $3\sigma$ uncertainties of equation of state of dark energy.}\label{fig:CPLeos}
\end{figure}

\begin{figure}
	\centering
	\includegraphics[scale=0.5]{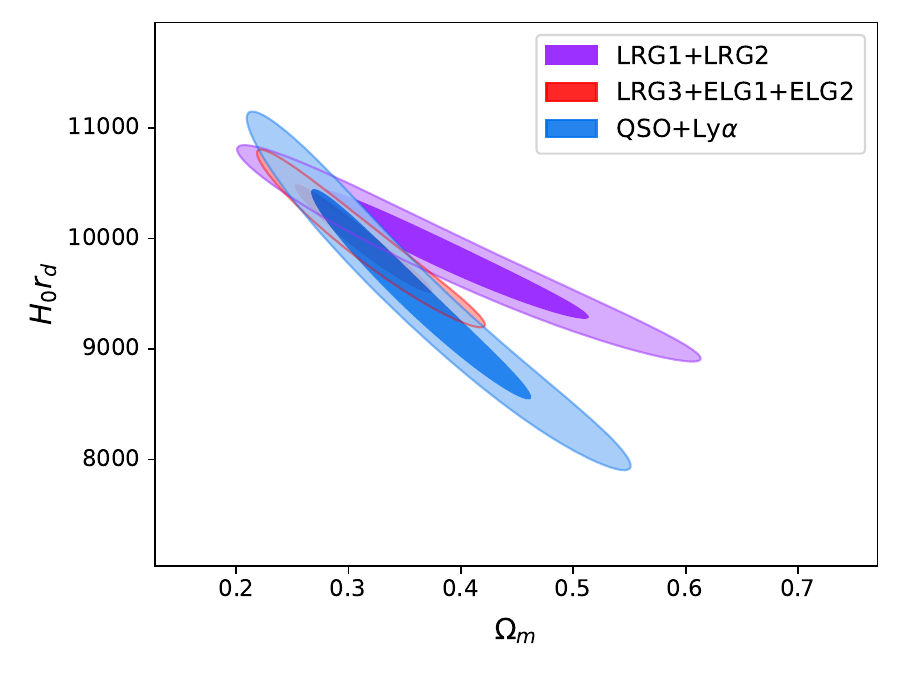}
	\caption{Two-dimensional posterior distributions of the parameter pair ($\Omega_m$, $H_0r_d$) from the latter three bins in our 4-bin model for DESI BAO measurements.}\label{fig:LCDMbin}
\end{figure}

\section{Data and methodology}
To study whether the cosmological analysis implemented by the DESI collaboration is self-consistent, we adopt the following observational datasets:

$\bullet$ CMB. Observations from the Planck satellite have very important meanings for cosmology and astrophysics \cite{Planck:2018vyg}. They have measured the matter components, the large scale structure of the universe and the topology. We employ the Planck 2018 high-$\ell$ \texttt{plik} temperature (TT) likelihood at multipoles  $30\leqslant\ell\leqslant2508$, polarization (EE) and their cross-correlation (TE) data at $30\leqslant\ell\leqslant1996$, and the low-$\ell$ TT \texttt{Commander} and \texttt{SimAll} EE likelihoods at $2\leqslant\ell\leqslant29$ \cite{Planck:2019nip}. We also take the Planck lensing likelihood \cite{Planck:2018lbu} from \texttt{SMICA} maps at $8\leqslant\ell \leqslant400$.

$\bullet$ BAO. BAO are very clean probes to explore the evolution of the universe over time, which are unaffected by the nonlinear physics at small scales. Measuring the positions of these oscillations in the matter power spectrum at different redshifts can provide strict constraints on the cosmic background expansion. We use 12 DESI BAO measurements specified in Ref.\cite{DESI:2024mwx}, including the BGS sample in the redshift range $0.1 < z < 0.4$, LRG samples in $0.4 < z < 0.6$ and $0.6 < z < 0.8$, combined LRG and ELG sample in $0.8 < z < 1.1$, ELG sample in $1.1 < z < 1.6$, quasar sample in $0.8 < z < 2.1$ and the Lyman-$\alpha$ Forest Sample in $1.77 < z < 4.16$ \cite{DESI:2024lzq,DESI:2024uvr}. 

$\bullet$ SN. Luminosity distances of type Ia SN are powerful distance indicators to probe the expansion history of the universe, especially, the equation of state of dark energy. We use the Pantheon+ SN sample \cite{Scolnic:2021amr}, which consists of 1701 light curves of 1550 spectroscopically confirmed type Ia SN coming from 18 different surveys.

To implement the numerical analysis, we use the publicly available Boltzmann solver \texttt{CAMB} \cite{Lewis:1999bs} and employ the Monte Carlo Markov Chain (MCMC) method to infer the posterior distributions of model parameters by using the public package \texttt{CosmoMC} \cite{Lewis:2002ah,Lewis:2013hha}. We choose the uniform priors for free parameters considered in this analysis: the baryon fraction $\Omega_bh^2 \in [0.005, 0.1]$, cold dark matter fraction CDM fraction $\Omega_ch^2 \in [0.001, 0.99]$, acoustic angular scale at the recombination epoch $100\theta_{MC} \in [0.5, 10]$, scalar spectral index $n_s \in [0.8, 1.2]$, amplitude of primordial power spectrum $\mathrm{ln}(10^{10}A_s) \in [2, 4]$, optical depth $\tau \in [0.01, 0.8]$, present-day equation of state of dark energy $\omega_0 \in [-4, 2]$ and  amplitude of dark energy evolution $\omega_a \in [-10, 5]$. Mote that we use the priors $\omega_0 \in [-4, 2]$ and $\omega_a \in [-10, 5]$ for the case of CMB alone or the data combination including CMB. In the DESI-only analysis, we use the sampler \texttt{emcee} \cite{ForemanMackey:2012ig} to perform the Bayesian analysis. For the CPL model, the priors we take are $\Omega_m \in [0.01, 0.9]$, $\omega_0 \in [-15, 20]$ and $\omega_a \in [-30, 10]$ and $2946 < H_0r_d < 14730$. To produce a matter-dominated era at high redshifts, we impose the condition $\omega_0 + \omega_a < 0$ in the MCMC analysis. Here the reason why we take such large prior ranges for the parameter pair ($\omega_0$, $\omega_a$) is that we need a large enough parameter space to completely present the constraining power of the DESI BAO measurements. We take the online package \texttt{Getdist} \cite{Lewis:2019xzd} analyze these MCMC chains. 

For simplicity, hereafter we use the DESI notation for each BAO measurement (see also Table 1 and Figure 2 in Ref.\cite{DESI:2024mwx}). Particularly, we denote the LRG data point at $z_{\rm eff}=0.51$ as LRG1.

\section{$\Lambda$CDM}
As is well known, at late times, $\Lambda$CDM has only one parameter, namely $\Omega_m$, which denotes present-day matter density fraction of the universe. The value of this parameter affects subtly the whole picture of modern cosmology. Planck-2018 CMB observations gives $\Omega_m=0.3153\pm0.0073$ \cite{Planck:2018vyg} at the $1\,\sigma$ CL, while the latest DESI BAO measurements provide the $1\,\sigma$ constraint $\Omega_m=0.295\pm0.015$ \cite{DESI:2024mwx}. The Pantheon+ with the SH0ES Cepheid host calibration gives $\Omega_m=0.334\pm0.018$ \cite{Scolnic:2021amr}. It is easy to find these three independent measurements of dark energy agrees well with each other at about $1\,\sigma$ CL. Therefore, so far, within the $\Lambda$CDM paradigm, Planck, DESI and Pantheon+ give basically the consistent prediction. 

Since two LRG data points at $z_{\rm eff}=0.51$ give a higher $\Omega_m$ in the case of DDE, we are interested in investigating their effect on $\Omega_m$ in $\Lambda$CDM. We obtain the constraint $\Omega_m=0.293\pm0.016$, which implies that LRG1 hardly affects the constraints on $\Lambda$CDM.

\section{CPL DDE}
CPL model is characterized by three free parameters, i.e., present-day equation of state of dark energy $\omega_0$, the amplitude of dark energy evolution with time $\omega_a$ and $\Omega_m$. One can easily find that CPL is a two-parameter extension to $\Lambda$CDM in the dark sector of the universe. In the above section, we gives an overview of the constraining results of $\Lambda$CDM. Similarly, in this section, we carry out the constraints on $\Omega_m$ for CPL.

In Fig.\ref{fig:CPLom}, it is easy to see that Planck, DESI and Pantheon+ gives consistent constraints at around $1\,\sigma$ CL, even though CMB alone give a lower peak value. This is because the errors of $\Omega_m$ are large for CPL. Specifically, Planck, DESI, Pantheon+ gives the $1\,\sigma$ constraints $\Omega_m = 0.381^{+0.063}_{-0.041}$, $\Omega_m = 0.365^{+0.15}_{-0.080}$ and $\Omega_m = 0.251^{+0.023}_{-0.11}$. Since the posterior density distribution of $\Omega_m$ from Planck is non-Gaussian, we also give the $2\,\sigma$ constraint $\Omega_m = 0.25^{+0.24}_{-0.12}$ as a comparison.

Furthermore, Ref.\cite{Colgain:2024xqj} gives the 4-parameter constraints on the CPL model using DESI alone. However, due to small prior ranges, a good constraint on two dark energy parameters $\omega_0$ and $\omega_a$ are not provided. We also notice that the DESI collaboration also take the small prior ranges $\omega_0 \in [-3, 1]$ and $\omega_a \in [-3, 2]$ to implement the constraints using their BAO data points, and obtain $\omega_0=-0.55^{+0.39}_{-0.21}$ and the $2\,\sigma$ upper limit $\omega_a<-1.32$ \cite{DESI:2024mwx}. We think the DESI's CPL constraint via its own data alone is incomplete and argue that the CPL parameter space should be sufficiently sampled so that we can obtain the reasonable constraint. Based on this concern, we impose large enough prior ranges for ($\omega_0$, $\omega_a$) and present our results in Tab.\ref{tab:CPL} and Fig.\ref{fig:CPLLRG}. We find that the $1\,\sigma$ constraint on today's equation of state of dark energy should be $\omega_0=1.04^{+0.91}_{-1.00}$, which prefers a positive value. Interestingly, the $2\,\sigma$ constraint $\omega_0=1.04^{+2.00}_{-1.90}$ gives a beyond $2\,\sigma$ evidence of quintessence dark energy. In light of the large parameter priors, we also obtain the compete constraint on the amplitude of evolution of dark energy $\omega_a=-7.4^{+3.8+6.8}_{-3.2-7.3}$ indicating a beyond $2\,\sigma$ preference of DDE. This consequence can be naturally explained by a quintessence scalar field at late times \cite{Ratra:1987rm,Wetterich:1987fm,Doran:2000jt,Caldwell:1997ii,Carroll:1998zi,Dvali:2001dd}. If discarding LRG1 in DESI, we find these two evidences reduce to about $1\,\sigma$ CL, although these 10 data points give compatible constraints with the whole DESI sample at $1\,\sigma$ CL. 

In Fig.\ref{fig:CPLLRG}, one can also easily observe that LRG1 provides a strong constraining power for CPL. Subsequently, we find an approximate fitting formula $3.62\times\omega_0+\omega_a=-3.73$ can well describe the posterior samples. This implies that DESI is actually measuring the linear combination $3.62\times\omega_0+\omega_a$ (see Fig.\ref{fig:CPLlinear}). Furthermore, a very interesting question that whether LRG1 affects the DDE preference from CMB+DESI+SN (hereafter CDS) emerges. In Fig.\ref{fig:CPLCDS}, we find that CDS with and without LRG1 give very consistent constraints indicating that LRG1 hardly affects the global fit. Especially, we obtain the $1\,\sigma$, $2\,\sigma$ and $3\,\sigma$ constraints $\omega_0=-0.707^{+0.089+0.18+0.24}_{-0.089-0.17-0.22}$ and $\omega_a=-1.09^{+0.38+0.67+0.82}_{-0.31-0.72-1.00}$ from CDS. Very interestingly, we find a beyond $3\,\sigma$ evidence of $\omega_0>-1$ and $\omega_a<0$, which is higher than $2.5\,\sigma$ from the analysis by the DESI collaboration \cite{DESI:2024mwx}. This enhancement is because we do not include the ACT DR6 lensing data \cite{ACT:2023kun} in this work. 

Moreover, we are interested in studying the effects of different probes on the $\omega_0$-$\omega_a$ plane. In Fig.\ref{fig:CPLw0wa}, we exhibit constraints on the parameter pair ($\omega_0$, $\omega_a$) from different probes and their combinations. It is easy to see that the preference of $\omega_0>-1$ and $\omega_a<0$ dominates the constrained parameter spaces from independent probes such as Planck, DESI or Pantheon+ within $1\,\sigma$ CL. Especially, the main part of ($\omega_0$, $\omega_a$) from DESI lies in the region of $\omega_0>-1$ and $\omega_a<0$. Even if no LRG1, this result still stands for DESI (see Fig.\ref{fig:CPLlinear}). In light of this, CMB+DESI can give a preference of DDE at $\sim2\,\sigma$ CL, while CMB+SN or DESI+SN gives a $\sim2\,\sigma$ detection of DDE with a slightly different degeneracy direction. Very interestingly, the addition of SN to CMB+DESI can, to a large extent, compresses the parameter space and leads to the above-mentioned beyond $3\,\sigma$ detection of DDE. Note that the addition of SN to CMB+DESI can reduce the allowed region of $\omega_0$-$\omega_a$ better than the addition of DESI to CMB+SN here. We argue that more significant preferences of DDE when combining CMB+DESI with Union3 or DESY5 SN samples are reasonable, because Pantheon+, Union3 and DESY5 share some same SN points and they have very similar degeneracy directions in the $\omega_0$-$\omega_a$ plane. 

An important issue is how dark energy evolves over redshift since we find a beyond $3\,\sigma$ preference of DDE. We present the dark energy evolution over redshift in Fig.\ref{fig:CPLeos} with $1\sigma$, $2\sigma$ and $3\sigma$ uncertainties. It is easy to see that CMB+DESI+SN prefers quintessence at beyond $3\,\sigma$ CL when $z\lesssim0.1$. Actually, this preference is very close to $4\,\sigma$ based on the constraint $\omega_0=-0.707^{+0.089+0.18+0.24}_{-0.089-0.17-0.22}$. One can also find that dark energy start to become phantom at the $2\,\sigma$ CL when $z\sim4$. These results lead to a remarkable phantom-crossing behavior. It is worth noting that all the DDE related results are obtained in the CPL model. 

An extended dark energy analysis can be found in \cite{Carloni:2024zpl,Cortes:2024lgw}. Especially, in Ref.\cite{Wang:2024hks}, five popular cosmological scenarios including inflation, modified gravity, annihilating dark matter, interacting dark energy and massive sterile neutrinos are studied carefully in light of the latest DESI BAO data..

\section{Thoughts for binning methods}
During the past 25 years, we always confront a cosmological model (e.g. $\Lambda$CDM) with observations in a global redshift bin. For example, here we consider DESI in a global redshift bin $z \in [0.1, 4.16]$. In order to study the evolution of a specific physical quantity over time, one usually divides the data into different subsets in different redshift bins. At first glance, this approach can give details of the universe at different epochs. However, it will enlarge the intrinsic fluctuations of data if considering a small redshift bin. In a simple system consisting of sources (e.g., galaxies), environments, detectors (e.g., DESI) and humans, the BAO data obtained is actually a distribution over redshift and many related elements that affect the real observations. If one just takes a small data subset in a small redshift bin into account, this subset may lose the globally average property of a physical quantity of interest from the whole sample, and may give a singular constraint. Therefore, we think an appropriate binning method for a specific dataset is needed if one wants to consider the details of a possible evolution. Although losing a small part of the information of the full distribution, a good binning can give evolution details that the whole sample can not do. On the flip side, one can not overestimate the impact of a constraining value from a small bin in cosmological analyses, even if effective correlations between different bins are considered. 

Along this logical line, we divide the DESI BAO measurements into four subsets: (i) BGS in $0.1 < z < 0.4$; (ii) LRG1+LRG2 in $0.4 < z < 0.8$; (iii) LRG3+ELG1+ELG2 in $0.8 < z < 1.6$; (iv) QSO in $0.8 < z < 2.1$ plus Ly$\alpha$ QSO in $1.77 < z < 4.16$ (hereafter QSO+Ly$\alpha$). We use the DESI's notation here (see Table 1 and Figure 2 in Ref.\cite{DESI:2024mwx}). Interestingly, LRG1+LRG2 gives the $1\,\sigma$ and $2\,\sigma$ constraints $\Omega_m = 0.391^{+0.072+0.170}_{-0.089-0.160}$, which is now more consistent with the global value $\Omega_m=0.295\pm0.015$, while LRG3+ELG1+ELG2 and QSO+Ly$\alpha$ give $\Omega_m = 0.313^{+0.037+0.082}_{-0.043-0.077}$ and $\Omega_m = 0.364^{+0.056+0.140}_{-0.073-0.120}$, respectively. This is as expected by our thoughts above. One can also see the latter three bins give consistent constraints in Fig.\ref{fig:LCDMbin}. Since the BGS sample just includes one data point and consequently can not provide a good constraint, we do not show its constraining result in this work (see Figure 2 in Ref.\cite{DESI:2024mwx} for details).

To go a further step to test our viewpoints on binning method, we also divide DESI BAO data into two subsets, i.e., BGS+LRG1+LRG2+LRG3+ELG1+ELG2 and QSO+Ly$\alpha$. We obtain the $1\,\sigma$ constraint $\Omega_m = 0.293\pm 0.023$ for the first bin, which is very consistent with the DESI's global result $\Omega_m=0.295\pm0.015$. Once again, a large bin will approximate the whole data distribution better than a small bin. The large biases from small bins are washed out when considering more subsets.

\section{Discussions and conclusions}
The first year of DESI BAO observations can help investigate the background dynamics of the universe and elucidate the nature of dark energy better than before. The DESI collaboration report the preferences of DDE at the $2.5\,\sigma$, $3.5\,\sigma$ and $3.9\,\sigma$ CL when combining DESI with Planck CMB and Pantheon+, Union3 and DESY5 SN samples, respectively. These joint constraints on DDE from three independent probes are fairly reasonable. However, we demonstrate that the DDE constraints implemented by the DESI collaboration may be insufficient or incomplete using their own BAO data. Using large enough prior ranges for $\omega_0$ and $\omega_a$, we obtain the constraints $\omega_0=1.04^{+0.91+2.00}_{-1.00-1.90}$ indicating a beyond $2\,\sigma$ preference of quintessence dark energy today and $\omega_a=-7.4^{+3.8+6.8}_{-3.2-7.3}$, which implies an evidence of evolving dark energy at beyond $2\,\sigma$ CL. Our results are different from $\omega_0=-0.55^{+0.39}_{-0.21}$ and the $2\,\sigma$ upper limit $\omega_a<-1.32$ reported by the DESI collaboration \cite{DESI:2024mwx}. This consequence can be naturally explained by a quintessence scalar field or any physical mechanism predicting a quintessence-like equation of state of dark energy at late times.

We also find that LRG1 hardly affect the constraints from the combined datasets CDS. Especially, we obtain the $1\,\sigma$, $2\,\sigma$ and $3\,\sigma$ constraints $\omega_0=-0.707^{+0.089+0.18+0.24}_{-0.089-0.17-0.22}$ and $\omega_a=-1.09^{+0.38+0.67+0.82}_{-0.31-0.72-1.00}$ from CDS. Very interestingly, we find a beyond $3\,\sigma$ evidence of $\omega_0>-1$ and $\omega_a<0$, which is higher than $2.5\,\sigma$ CL from the DESI's analysis \cite{DESI:2024mwx}. This enhancement is because we do not include the ACT DR6 lensing data in our analyses. Actually, this preference of quintessence-like DDE reaches about $4\,\sigma$ CL (see Fig.\ref{fig:CPLeos}).

Furthermore, we notice that dark energy start to become phantom at the $2\,\sigma$ CL when $z\sim4$. At the same time, the constraint from CDS leads to a significant phantom-crossing behavior. It is noteworthy that all these results are obtained in the CPL model. We argue that there may be a model dependence for the constrained dark energy behavior from the CDS observations.  

Finally, we make some simple comments on the widely used binning method in cosmological analyses. We argue that an appropriate binning method for a specific dataset is needed if one wants to study the evolution of a cosmological quantity.  A good binning approach can provide more evolution details that the whole sample can not obtain. Nonetheless, the price is losing a small part of information of the data distribution. On the contrary, one can not overestimate the impact of the result from a small bin in cosmological analyses, since it may have a large bias induced by the intrinsic fluctuation of data.
We have also used the DESI data to demonstrate the correctness of our viewpoints of binning methods.

In the forthcoming years, the full DESI survey will help unveil the nature of dark energy better in synergy with CMB and SN experiments.

\section{Acknowledgements}
DW thanks Olga Mena and Eleonora Di Valentino for useful communications and discussions on cosmological data analyses.
DW is supported by the CDEIGENT Project of Consejo Superior de Investigaciones Científicas (CSIC).

\end{document}